\documentstyle[aps,preprint,12pt]{revtex}
\begin{document}
\title{Effects of Mobility of Small Islands on Growth \\
in Molecular Beam Epitaxy}
\author{
Itay Furman \cite{itay}
and 
Ofer Biham \cite{ofer}
}
\address{
Racah Institute of Physics,
The Hebrew University,
Jerusalem 91904,
Israel
}
\maketitle

\begin{abstract}
The effects of mobility of small islands on island growth
in molecular beam epitaxy are studied. It is shown that 
mobility of small islands affects both the scaling and morphology of 
islands during growth.
Three microscopic models are considered, in which the critical
island sizes are $i^*=1,2$ and $3$ 
(such that islands of size $s \le i^{\ast}$ are {\it mobile} while
islands of size $s \ge i^{\ast}+1$ are {\it immobile}). 
As $i^*$ increases, islands become more compact,
while the exponent $\gamma$ which relates the island density to
deposition rate increases.
The morphological changes are quantified by
using fractal analysis. 
It is shown that the fractal dimensions are
rather insensitive to changes in $i^{\ast}$.
However, the prefactors provide a quantitative 
measure of the changing morphologies. 

\end{abstract}

\pacs{68.35.Fx,61.43.Hv,68.55.-a,82.20.Mj,82.20.Wt}

\newpage

\section{Introduction}
	\label{intro.}

The growth of thin films in molecular beam epitaxy involves
atom diffusion on the surface and nucleation of islands,
followed by aggregation and coalescence.
The resulting morphology is found to depend on the detailed 
nature of microscopic diffusion processes on the substrate,
as well as the deposition rate and substrate temperature.
Scanning tunneling microscopy (STM) has revealed
a variety of morphologies in the submonolayer regime.
Compact islands of nearly square shape 
have been observed for homoepitaxial growth
on FCC(001) substrates such as Ni(001) and Cu(001)
\cite{kopatzki93,gunther94}.
Fractal-like islands which resemble 
diffusion limited aggregation
(DLA) 
\cite{witten81}
clusters have been observed in systems such as 
Au on Ru(0001)
\cite{hwang91}
while Pt on Pt(111) was 
found to exhibit both fractal-like shapes
and more compact, nearly triangular shapes
\cite{michely93,bott92}.
Experimental studies and theoretical 
work using Monte Carlo simulations
have shown that the submonolayer 
morphology depends on the rates of various 
diffusion processes on the substrate.
In particular, systems which
exhibit high atom mobility along island edges 
(typically on square substrates)
tend to form  compact islands.
Low edge mobility gives rise to islands with DLA-like shapes
(typically on hexagonal substrates).
These morphologies are important beyond the submonolayer since
they affect the multilayer growth mode of the film.
The island morphology also affects the Schwoebel barrier
\cite{schwoebel69}
for an atom deposited on top of an island to hop down
the step.
When this barrier is large islands tend to nucleate 
on top of islands resulting in three 
dimensional growth mode while a small barrier gives rise to 
layer by layer growth.

Scaling properties of island growth have been studied
experimentally using statistical analysis of STM data
and helium beam scattering
\cite{mo91,zuo91,ernst92,li93,stroscio94,zuo94,durr95}. 
Theoretical studies have
applied rate equations 
\cite{stoyanov81,venables84,villain92,bartelt92,tang93,zangwill93,biham95}
as well as Monte Carlo (MC) simulations
\cite{voter86,clarke88,bartelt93,smilauer93,barkema94,bales94,ratsch94,zhang94
,jacobsen94,schroeder95,amar95,bales95,linderoth96,biham96}. 
It was found that for a broad class of systems
a short transient regime is followed by a quasi-steady state
where a scaling relation of the form
\begin{equation}
N \sim \left( {F \over 4 h} \right)^{\gamma} 
\end{equation}
applies,
where $N$ is the island density, 
$F$ is the flux or deposition rate and
$h$ is the adatom diffusion coefficient
(note that the hopping rate for an isolated adatom 
is $4 h$ due to the four possible directions for hopping).
The exponent $\gamma$ depends on microscopic properties such
as the stability 
\cite{amar95}
and mobility 
\cite{villain92,bartelt96}
of small islands,
isotropic vs. anisotropic diffusion
\cite{gunther94,mo91,linderoth96}
and the existence of magic islands
(namely, islands which are stable 
while larger islands are unstable)
\cite{schroeder95}.
It was found that for systems in which the smallest stable island
is of size $i^\ast + 1$ (where islands of size $s \le i^\ast$ dissociate),
in the asymptotic limit of slow deposition rate,
for isotropic diffusion,
$\gamma = i^\ast / (i^\ast + 2)$.
Experimentally, it was found that $\gamma$ may depend on temperature
\cite{ernst92,zuo94}. 
For example, for FCC(001) surfaces a transition from $i^\ast = 1$
to $i^\ast = 3$ due to temperature increase was characterized,
where the transition temperature is determined by the ratio between
the aggregation and dissociation rates of dimers and trimers
\cite{bartelt_istar}.
The island size distribution was also studied
and found to depend on $i^\ast$. 
These studies revealed that scaling 
properties can be used to identify microscopic processes
at the atomic level and to estimate various activation energies 
which are difficult to measure.
Experimental measurements of some diffusion processes at
the atomic level are possible using field ion microscopy.
This technique was used to identify diffusion modes for small
islands such as dimers and trimers 
on FCC(001) metal surfaces
and to measure their diffusion
coefficients
\cite{kellogg91}.

In this paper we consider the relation between the scaling and 
morphology of the growing islands and trace them to properties
of the microscopic model. In particular, we examine the effect
of the {\it mobility} of small 
islands on island morphology and scaling.
To this end we introduce a class of three microscopic models 
on a square substrate which differ in their diffusion properties. 
In model I only single atoms (monomers) 
are mobile [which means that the 
energy barrier for breaking a nearest neighbor (nn) or 
next-nearest-neighbor (nnn) bond is practically infinite]. 
In model II both monomers
and dimers (clusters of two atoms)
are mobile while in model III monomers, dimers and trimers
(clusters of three atoms) are mobile.  
An island is defined as a cluster of adatoms connected by nn or
nnn bonds. 
The island size $s$ is the number of atoms in the island. 
In this paper we do not consider island instability
but only island mobility.
Therefore we define the critical island size $i^{\ast}$ as the size for
which all islands of size $s \le i^{\ast}$ are mobile while
islands of size $s \ge i^{\ast}+1$ are immobile. Thus, 
$i^{\ast}=1,2$ and $3$ for models I, II and III respectively.
Note that in the models considered here islands are stable even for
$s \le i^{\ast}$ in the sense that there is no detachment of atoms
from islands. 
However, due to their mobility these small 
islands tend to collide and
merge into other islands.

The models we consider involve only local 
(nn and nnn)
interactions such that 
the hopping rate for an atom 
in each of the four possible directions
is determined by the occupancy 
in a neighborhood of $3 \times 3$ sites around it.
We show that in this type of models there is
a correlation between mobility of small islands and edge mobility.
In model I, once an adatom attaches to an island edge as a
nn it cannot move and therefore edge 
mobility is completely suppressed.
Model II allows very limited edge mobility 
while model III allows more moves including hopping around a corner
of an island. 
As the edge mobility increases islands become more compact because
it tends to suppress narrow fingers and allows the atoms to find
more stable positions with more nearest and next nearest neighbors.

The correlation between mobility of small islands and edge 
mobility gives rise to a relation between the scaling properties
determined by the critical island size and the morphology which
depends on the edge mobility. To quantify this relation we first
obtain the scaling exponent $\gamma$ for the three models
using both rate equations and MC simulations. 
We then examine the island morphology for the three models and
apply a fractal analysis using the box counting algorithm 
and mass dimension evaluation
\cite{falconer90}.
The box counting function, in the scaling regime, can be described
by
\begin{equation}
N_{B}(\ell) = A_B \cdot {\ell}^{-D_B}
\end{equation}
where $\ell$ is the box size, $N_{B}(\ell)$ is the number of boxes which
contain at least one atom, 
$D_B$ is the box counting fractal dimension and 
$A_B$ is the prefactor. 
We find that the fractal dimension
$D_B$ is rather insensitive to the differences between the models.
The prefactor $A_B$, however, 
provides a quantitative measure of the different morphologies.
Similar conclusions are obtained for the mass dimension
$D_M$ and its prefactor $A_M$.

The paper is organized as follows.
The models are introduced in Section II. 
Scaling properties of island growth and their
dependence on the microscopic model are presented in Section
III. The morphologies are examined in Section IV, followed
by discussion in Section V and a
summary in Section VI.

\section{models}
\label{sec:models}
To study the scaling and morphology during island growth
we introduce three
models of diffusion on the square lattice. 
Submonolayer thin film growth during deposition in molecular beam
epitaxy is then studied using kinetic MC simulations
\cite{fichthorn91}.
In these simulations atoms are deposited randomly on the square
substrate at a rate $F$ 
[given in monolayers (ML) per second]
and hop according to the microscopic 
model.
The 
hopping rate
$h$ 
(in units of hops per second)
for a given atom 
to each unoccupied nn site
is given by
\begin{equation}
h=\nu \cdot \exp (-E_B/k_B T),
\label{hop_rate}
\end{equation}
where
$\nu=10^{12}$ $s^{-1}$ is the attempt frequency, 
$E_B$ is the energy barrier,
$k_B$ is the Boltzmann factor and $T$ is the temperature.
The coverage after time $t$ is then
$\theta = F \! \cdot \! t$ (in ML).

In our MC simulations moves are selected randomly 
from the list of all possible moves at the given 
time with the appropriate weights.
The time is  then advanced according to the inverse of the sum of
all rates.
In the models used here the energy barrier 
$E_B$
for hopping is determined by
the local environment in a $3 \times 3$ square around the hopping
atom (Fig. 1), where the occupancy of seven adjacent sites,
$k=0,\dots,6$ is taken into account. 
Each one of these sites can be either occupied ($S_k = 1$) 
or vacant ($S_k=0$) giving rise to $2^7=128$ barriers.
To index them we use the binary representation
where the energy barriers are
$E_B^n$, $n=0,\dots,127$ and 
$n=\sum_{k=0}^6 S_k \! \cdot \! 2^k$. 
Using this indexing we introduce three models of nearest neighbor
hopping on the surface
(Fig. 2).
In model I only monomers are mobile, 
in model II dimers are also mobile
and model III includes mobility of trimers as well.
To simplify the analysis, our models include only one hopping rate,
obtained by assigning the same activation energy 
($E_B=0.5$ $eV$)
to all the
allowed moves and a very high 
(practically infinite) energy barrier for all
moves which are not allowed in the given model.
The models are minimal in the sense that they include only 
the minimal set of moves required to achieve the 
specified island mobility. Some more moves can be added without
modifying $i^{\ast}$ (see Section V).
Note that in our models there is no additional barrier for 
hopping down a step, which is justified for small islands and 
especially for fractal-like ones
\cite{jacobsen94}.
Atoms deposited on top of an island hop until they hit the edge and
hop down to be incorporated into the island.
Nucleation of a second layer is thus suppressed.
The models are used to
perform a systematic study of the effects of mobility of small islands 
on the scaling and morphology of the growing islands.
The three models are introduced below.

\subsection{Model I}

In this model only monomers can move 
[Fig. \ref{moves} (a)].
Atoms cannot move away from 
nn or nnn atoms while they can move
towards a nnn atom making it a nn (moves $4$, $64$ and $68$
in Fig.
\ref{moves}).
The latter  moves are included to enhance adatom association and the
creation of islands.
(Note that in this model none of the allowed moves involve bond
breaking.)

\subsection{Model II}

Here both monomers and dimers are mobile.
In addition to the moves included in model I two new moves are added
[Figure 
\ref{moves} (b)].
Dimer mobility is now possible via a combination of a bond-breaking
(moves $2$ or $32$) and a bond building move (moves $4$ or $64$).
These additional moves introduce
more channels for island rearrangements which
lead to a more compact island shape
[Fig. \ref{addmoves} (a)].

\subsection{Model III}

In this model trimers are mobile, too.
Four moves are added to those present in model II 
(moves $3$, $6$, $48$ and $96$).
which are required for trimer mobility
[Fig. 2 (c)].
They also enhance the mobility of atoms on island edges, and
in particular allow for an edge adatom to move around a corner
[Fig. \ref{addmoves} (b)].
Note, however that edge mobility in this model is still highly
limited. For example, an atom adjacent to a straight long edge is 
immobile even in Model III.

\subsection{Diffusion Coefficients}

To confirm that 
the critical island size is indeed $i^{\ast}=1,2,3$ for models
I, II and III respectively,
and to obtain 
the diffusion coefficients of mobile islands in each model
we have done simulations of single cluster diffusion.
To obtain the statistics required for a precise 
determination of the diffusion
coefficients we performed 1000 runs
for monomers, dimers, trimers and tetramers
(islands of four atoms) in each of the three models. 
Each run was carried out to time equal to $0.6$ seconds, which is
about $200$ times larger than the time scale for hopping
at the given temperature ($T=250 K$). 
The diffusion coefficients  
were obtained from the relation
$\langle r^2 \rangle = 4 h_s \! \cdot \! t$, $s=1,2,3,4$
where $r$ is the distance between the initial
position of the center of mass of the
cluster and its position after 
time $t$, $h_s$ is the diffusion coefficient
for a cluster of size $s$ and 
$\langle ... \rangle$ represents an average over the 1000 runs.
The diffusion coefficients 
for monomers, dimers, trimers and tetramers
in each of the three models 
are shown in
Table
\ref{tab:hopping}
\cite{diffusion_coef}.
Our expectations are confirmed,
namely, that in model I only
monomers are mobile, in model II dimers are mobile 
as well and in model III
also trimers are mobile.

\section{scaling properties}
	\label{sec:scaling}

The submonolayer growth is typically divided into three
stages - the early stage is dominated by island nucleation
followed by an aggregation-dominated 
stage until coalescence sets in.
In the aggregation stage the density
of stable islands $N$, exhibits
power law behavior as a function of the ratio between 
the deposition rate and
the adatom hopping rate 
of the form
$N \sim (F/4 h)^{\gamma}$.
The exponent $\gamma$ is determined by the microscopic 
processes that are activated on the surface during growth.
In case where all clusters of size
$s \le i^{\ast}$
are mobile,
while larger clusters are immobile, 
the asymptotic value of 
$\gamma$ 
in the limit  
where $F/h \rightarrow 0$ 
is given by
\cite{villain92}:
\begin{equation}
\gamma = \frac{i^\ast}{2 i^\ast + 1}.
\label{eq:gamma}
\end{equation}
This result is exact if all mobile islands have the same diffusion
coefficient, namely $h_s = h$, $s = 1,\ldots,i^\ast$.
Still, it is a good approximation
if all the diffusion coefficients are of the same order of magnitude
and the ratios between them are independent of the temperature,
as is the case for the models studied in this paper.

To study the scaling properties of island growth
with mobility of small islands, 
for experimentally relevant deposition rates,
we introduce a set of rate 
equations which describe the time evolution of the densities
of mobile and immobile islands.   
In these equations, islands including $s$ atoms are mobile
for 
$1 \leq s \leq i^{\ast}$ with diffusion coefficient $h_s$
and immobile
for
$s \geq i^{\ast}+1$. 
The density of mobile islands of size $s$ is given by $x_s$,
$1 \leq s \leq i^{\ast}$, while the total density of immobile
islands is given by $N$.
Both densities are normalized per lattice site
and the lattice constant is taken to be $1$.
The rate equations take the form
\begin{mathletters}
\label{set:mobil_f}
\begin{eqnarray}
	\dot{x}_1 & = & F - \sum_{i=1}^{i^{\ast}} 4 (h_1 + h_i) x_1 x_i  
		- 4 h_1 x_1 N
			\label{set:mobil_f-adatoms}
	\\
	\dot{x}_s & = &  \sum_{i=1}^{i^{\ast}} 2 (h_i + h_{s-i}) 
		x_i x_{s-i} - \sum_{i=1}^{i^{\ast}} 4 (h_i + h_s)
		x_i x_s  - 4 h_s x_s N
		\;\;\;\;\;\;\;\;\;\;\;\;
		2 \leq s \leq i^{\ast}
			\label{set:mobil_f-clusters}
	\\
	\dot{N} & = &  \sum_{i=1}^{i^{\ast}} 
                                       \sum_{\  j=i^{\ast}+1-i}^{i^{\ast}}
		2 (h_i + h_j) x_i x_j,
			\label{set:mobil_f-islands}
\end{eqnarray}
\end{mathletters}
where the density of immobile islands $N$, is given by 
\begin{equation}
\label{def:N}
  N  \equiv  \sum_{s=i^{\ast}+1}^{\infty} x_s		.
\end{equation}
The first term in Eq.\ 
(\ref{set:mobil_f-adatoms})
describes the deposition of new atoms, while the first term in 
(\ref{set:mobil_f-clusters})
describes the building of mobile 
islands by merging of two mobile islands of smaller sizes.
The second and third terms in Eqs.\
(\ref{set:mobil_f-adatoms})
and
(\ref{set:mobil_f-clusters})
describe the reduction in the number of mobile islands of size $s$ due to
their collision with other mobile and immobile islands, respectively.
The third equation describes the rate of nucleation of immobile islands
due to collision of two mobile islands.

A rate equation representation of 
models I, II and III 
can be obtained from 
Eqs. 
(\ref{set:mobil_f})
when
$i^{\ast}=1, 2$ and $3$ respectively
and with the diffusion coefficients $h_i$ given in
Table I.
Note that the rate equations provide a mean field description
ignoring spatial correlations. Moreover, since we concentrate
here on the effects of mobility of small islands, other processes
such as atom detachment from islands, atoms deposited on top of
islands and coalescence are not included in the equations. 
Also, for simplicity, the capture number is taken as a constant.
In general, it is found to depend on both the island 
size and the coverage
\cite{bales94}.

We have examined
Eqs.
(\ref{set:mobil_f})
in the aggregation regime 
using asymptotic analysis
of the type described in Ref. 
\cite{tang93}.
In this analysis Eqs.
(\ref{set:mobil_f})
are written in dimensionless form using
$\hat x_s = \ell_1^2 \! \cdot \! x_s$, 
$\hat N = \ell_1^2 \! \cdot \! N$ 
and 
$\hat t = t/t_1$
where
$\ell_1 = (4h/F)^{1 \over 4}$
and 
$t_1 = (4hF)^{- {1 \over 2}}$.
We then consider the aggregation stage
($\hat t \gg 1$)
assuming that
$\hat N \gg \hat x_1 \gg ... \gg \hat x_i^{\ast}$.
Since at this stage the density of mobile islands
is approximately constant we solve 
Eqs.
(\ref{set:mobil_f})
in its dimensionless form for
$\dot{\hat x_s} = 0$, $1 \le s \le i^{\ast}$.
This is done using the ansatz that
$\hat x_s \sim {\hat t}^{\alpha \cdot s + \beta}$
and 
$\hat N \sim {\hat t}^{\delta}$,
where $\alpha$, $\beta$ and $\delta$ are constants
to be determined from the solution.
Solving for the leading order in each equation
we find scaling relations for the 
densities of immobile and mobile islands as a function of 
$F/4h$ and $\theta$
\cite{oldscalingrelation}:
\begin{mathletters}
\label{set:mobil-scaling}
\begin{eqnarray}
  x_s  & \sim & \left({F /4h} \right)^{ \frac{i^{\ast}+s}{2i^{\ast}+1} }  
 \cdot \theta^{ -\frac{2s-1}{2i^{\ast}+1} }
		\;\;\;\;\;\;\;\;\;\;\;\;
		s = 1, 2, \ldots, i^{\ast}
			\label{set:mobil-cluster_scaling}
  \\
  N  & \sim & \left({F /4 h} \right)^{ \frac{i^{\ast}}{2i^{\ast}+1} }  
\cdot \theta^\frac{1}{2i^{\ast}+1}.
			\label{set:mobil-island_scaling}
\end{eqnarray}
\end{mathletters}
For models I, II and III
we find $\gamma = 1/3, \; 2/5$ and $3/7$, respectively.
It is important to note that these results are only
asymptotically exact
in the limit $F/4h \rightarrow 0$.
Numerical integration of the rate equations 
shows slow convergence to
these results 
as the deposition rate is lowered.
In Table
\ref{tab:scaling}
we present the  values for $\gamma$ obtained from numerical integration
of Eq. 
(\ref{set:mobil_f})
with diffusion coefficients $h_i$, taken from Table
\ref{tab:hopping},
together with the asymptotic values.

To complement these results we have also examined the scaling
of island density $N$ vs. deposition rate $F$, using 
MC simulations
(Fig. \ref{scaling}).
The island density obtained in these simulations for the three models,
as a function of deposition rate, is shown in 
Fig. 
\ref{scaling} 
for coverage of $\theta=0.2$
and substrate temperature $T=250 K$.
Each data point in this figure represents an average over 100 runs and
the lattice size is $250 \times 250$ in all the runs.
The values of $\gamma$ for the three models,
obtained from the MC results in Fig. 
\ref{scaling}
are summarized in Table
\ref{tab:scaling}.
It is observed that $\gamma$ increases as $i^{\ast}$
is increased and this trend appears in all
three columns of Table 
\ref{tab:scaling}
representing the asymptotic result, rate
equations and MC simulations.
This can be intuitively understood from the following qualitative
argument.
In the limit of very fast deposition rate (say $F/4h \cong 1$)
the effect of island mobility is negligible and the 
island density is very large and nearly independent of 
$i^{\ast}$.
As $F/4h$ is decreased the mobility of single atoms and
islands of size $s \le i^{\ast}$ gives rise to nucleation
of fewer and larger islands.
This process is enhanced as $i^{\ast}$ increases,
since it adds more channels for collision and merging
of islands. 
This results in a faster decrease of $N$ as a function
of $F/4h$.

The values of $\gamma$ for rate equation integration at a finite rate
are found to be lower than the asymptotic value. The $\gamma$ values
for MC simulations are higher than the corresponding rate equation
values which can be attributed to spatial correlations and coalescence
which are not taken into account in the rate equations. 
Wolf 
\cite{schroeder95}
proposed 
an alternative formula for the 
exponent 
$\gamma$ which takes into account the fractal dimension of the
islands denoted by $D_f$.
Adapted to our case it takes the form
\begin{equation}
\label{eq:fract_gamma}
\gamma = {2 i^{\ast} \over 4 i^{\ast} + D_f} 
\end{equation}
which coincides with 
Eq. 
(\ref{eq:gamma})
for compact islands ($D_f=2$).
Note that Eq.\
(\ref{eq:fract_gamma})
implies that $\gamma$ for fractal islands
is larger than for compact islands with the same $i^\ast$,
giving rise to lower densities of islands
\cite{schroeder95}.
This is reasonable since fractal islands have
a larger effective area per given mass.
Thus, they inhibit more efficiently new nucleation events,
which results in a smaller island density, $N$.
Since the numerical simulations generate fractal-like islands
while the rate equations assume point-like (and thus compact)
islands this might provide further explanation for tendency of
the MC exponents to be larger than the exponents obtained from
numerical integration of the rate equations.

The scaling properties of the island size distribution have
been studied both experimentally 
\cite{stroscio94}
and theoretically
\cite{bartelt92,bartelt93,schroeder95,amar95}.
These studies indicated that the island size distribution
strongly depends on the stability and mobility of small islands
and is modified in the case of magic islands
\cite{schroeder95}.
The scaled island size distributions for models I, II and III
are presented in Fig. 5(a-c).
Here $\bar s$ is the average island size.
For all three models the deposition rates are
$F=10^{-3}$ $ML/s$ ($\bigcirc$),
$F=10^{-4}$ $ML/s$ 
(\vbox{\hrule width 8pt height 8pt})
and
$F=10^{-5}$ $ML/s$ ($\times$).
We observe that as $i^{\ast}$ 
increases the peaks rise more slowly on the left hand size
(small $s/\bar s$) due to the depletion of the mobile islands.
They also fall off more sharply on the right hand side
and thus become narrower.
This trend is qualitatively similar to previous results for
the case where small islands are unstable.
Note that the peak height increases considerably as $F$ 
decreases.
This may be due to coalescence which is found to become more
pronounce as the deposition rate decreases.
Coalescence causes $\bar s$ to increase, pushing up the
scaled island size distribution 
which includes the factor
${\bar s}^2/\theta$.

\section{Morphology}
	\label{sec:morphology}

To examine the relation between 
mobility of small islands and island morphology 
we have performed extensive MC simulations of 
island growth using
the 
three models described above.
The morphology of the growing islands is shown in Fig.
\ref{snaps}
for deposition rate $F=10^{-6}$, 
coverage $\theta=0.2$
and $T=250 K$.
The morphology obtained for model I 
[Fig.
\ref{snaps} (a)]
best resembles the shape of small 
DLA clusters grown on the square lattice,
due to the suppressed edge mobility.
The islands of models II and III 
[Fig. 
\ref{snaps} (b) and (c), respectively]
exhibit wider arms and have more compact shapes
due to edge mobility associated with the increased $i^\ast$.
To quantify these observations we have performed fractal analysis
of the island morphology using the box counting algorithm 
and mass dimension evaluation
for the three models.
In the box counting technique one divides the lattice into boxes
of linear size $\ell$ 
and counts the
number of boxes $\langle N_{B}(\ell) \rangle$ 
(averaged over a series or runs) 
which intersect the set of islands.
The box counting dimension $D_B$ is then given by
\begin{equation}
\log_{10} \langle N_{B}(\ell) \rangle = 
\log_{10}  A_B - D_B \cdot \log_{10} \ell
\label{def:boxc_dim}
\end{equation}
where $A_B$ is a prefactor.
The box counting function $N_{B}(\ell)$ for the three models
is shown in Fig. 
\ref{boxc}
on a $\log-\log$ scale.
The fractal dimensions are presented in Table
\ref{tab:d_b}
and the prefactors in Table
\ref{tab:A_b}.
It is shown that although the three models generate different looking
morphologies the box counting dimension 
$D_B$ 
for a given deposition
rate and coverage is practically independent of the model.
The box counting dimensions are 
found to be 
typically lower than the DLA dimension 
$D_{DLA} \cong 1.72$
and
increase as the deposition rate decreases.
A significant difference between the models is reflected in
the coefficient $A_B$. 
It is shown 
(Table \ref{tab:A_b})
that as 
$i^{\ast}$ increases and
islands become more compact, 
$A_B$ decreases by almost a factor of two.
It also decreases when deposition rate is increased (which also results
in more compact islands).
The behavior of $A_B$ can be understood from the fact that as islands
become more compact each occupied box tends to include more atoms
(and still it is counted only once).
Since the coverage is maintained the number of occupied boxes must
decrease.

Unlike the box counting analysis which is done on the entire system,
the mass dimension analysis is done on each island separately. In this
analysis one finds the center of mass of the island and measures the
total mass of the atoms $M(r)$
bounded by a circle of radius $r$ around it as 
a function of $r$.
The mass dimension $D_M$ is then obtained from 
(Fig. \ref{massd})
\begin{equation}
\log_{10} \langle M(r) \rangle = \log_{10} A_M + D_M \cdot \log_{10} r.
\label{def:mass_dim}
\end{equation}
where the average is over a large number of islands.
The values of the mass dimension $D_M$
obtained for the three models are 
summarized in Table
\ref{tab:d_m}
and the prefactors 
$A_M$
in Table 
\ref{tab:A_m}.
The mass dimension is found to be in the range between $1.83$ and
$1.91$, significantly larger than the DLA dimension.
This larger dimension reflects the enhanced compactness of the
islands, however its dependence on the model is rather weak. 
The morphological differences are
strongly reflected in the coefficient $A_M$  which increases 
as $i^{\ast}$ is increased.
This trend of $A_M$ results from the fact that as the islands become
more compact the mass included in a circle of radius $r$ around their
center must increase.
We conclude that the scaling properties reflected
in $D_B$ and $D_M$ are weakly dependent on the models,
while the prefactors $A_B$ and $A_M$ are strongly model dependent.
In all the three models we considered, edge mobility is limited
and allows only local rearrangements of the structure.
The prefactors (which are related to the curve intercept, {\it i.e.},
to small scales) are sensitive to such changes.
In contrast, the fractal dimensions which are determined over
a range of scales are rather insensitive to the small scale behavior.

Note that in model I the diffusion is similar to the DLA model.
However, unlike DLA atoms are added at a finite
rate nucleating a finite density of islands.
Also, these atoms are deposited randomly from above rather than
from boundaries located far away from the cluster as in DLA. 
Therefore, some of the atoms fall between 
(or on top of)
the arms of the island.
For these atoms the aggregation process is not
diffusion limited.
The DLA limit can be approached only when both
$F/4h \rightarrow 0$ and 
$\theta \rightarrow 0$.
The first limit is required in order to keep islands
far away from each other. 
The second limit is required in order to ensure that
the islands are very small compared to the area from 
which they draw atoms.

\section{Discussion}

The fractal analysis can provide useful information about 
processes and rates in island growth systems. It is 
accessible experimentally and can be done
using
either STM data 
\cite{hwang91}
or helium beam scattering
\cite{hamburger95}.
In the latter case,
one possibility is to obtain the fractal dimension of contours
of constant electron density of the monolayer from measurements
of the specular peak intensity as a function of helium 
incidence energy.

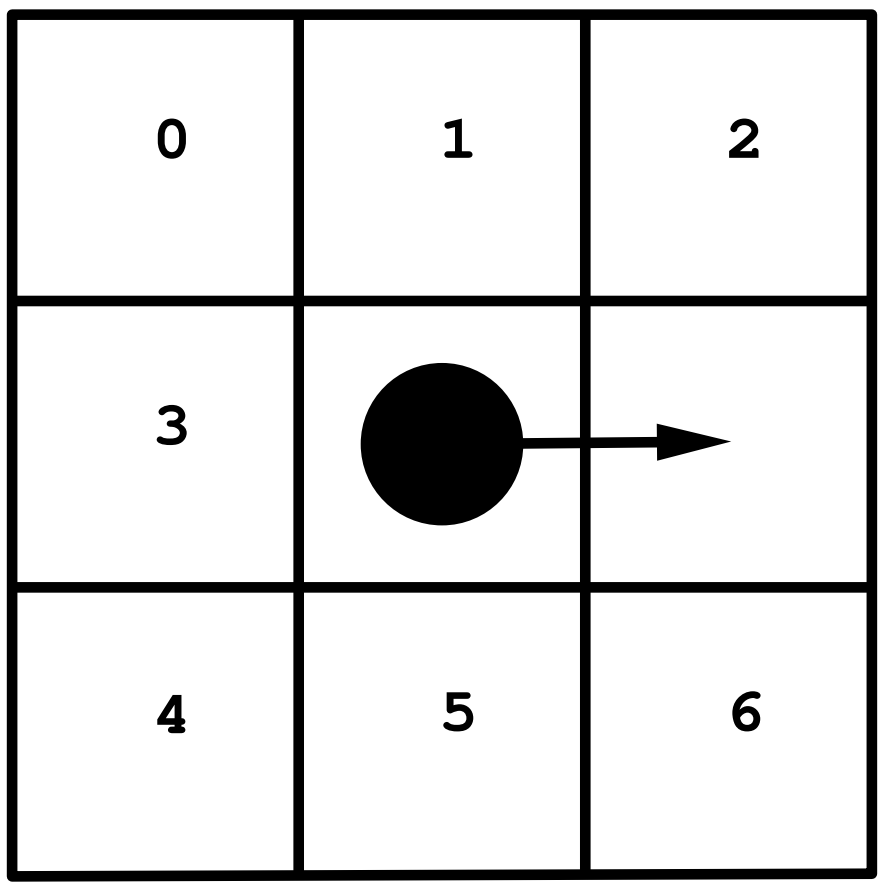
\begin{figure}
\caption{
The local environment of an hopping atom. 
Each one of the seven adjacent sites 
$k=0,\dots,6$
can be either 
occupied ($S_k=1$) or 
unoccupied ($S_k=0$),
giving rise to $2^7=128$ local environments
with activation energies
$E_B^n$, $n=0,\dots,127$ where 
$n=\sum_{k=0}^6 S_k \cdot 2^k$.
}
\label{local} 
\end{figure}

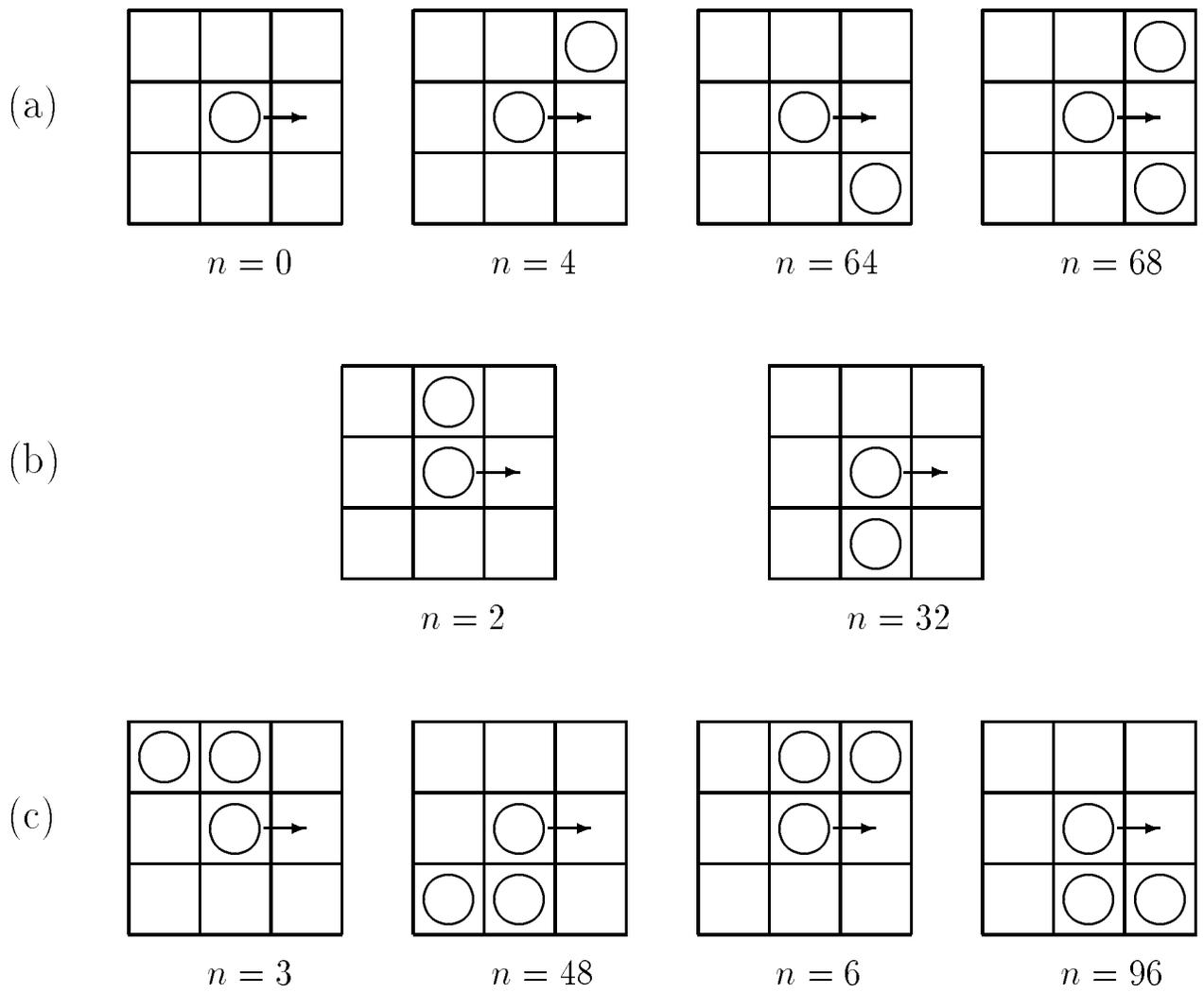
\begin{figure}
\caption{
The moves included in the three models. Model I includes only
the moves in (a). Model II includes the moves in (a) and (b)
while model III includes all the moves in (a), (b) and (c).
All the allowed moves have the same activation energy $E_B$,
while the activation energy for all other moves is practically
infinite.
}
\label{moves} 
\end{figure}

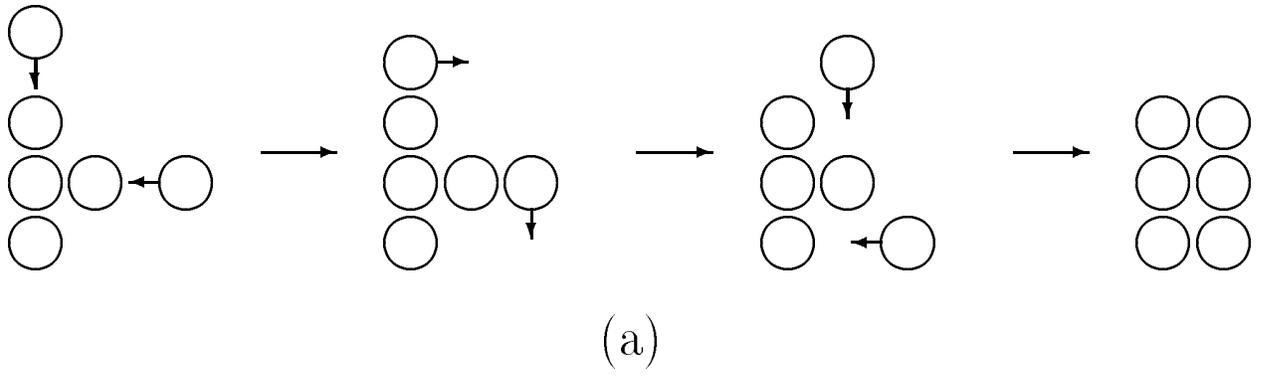
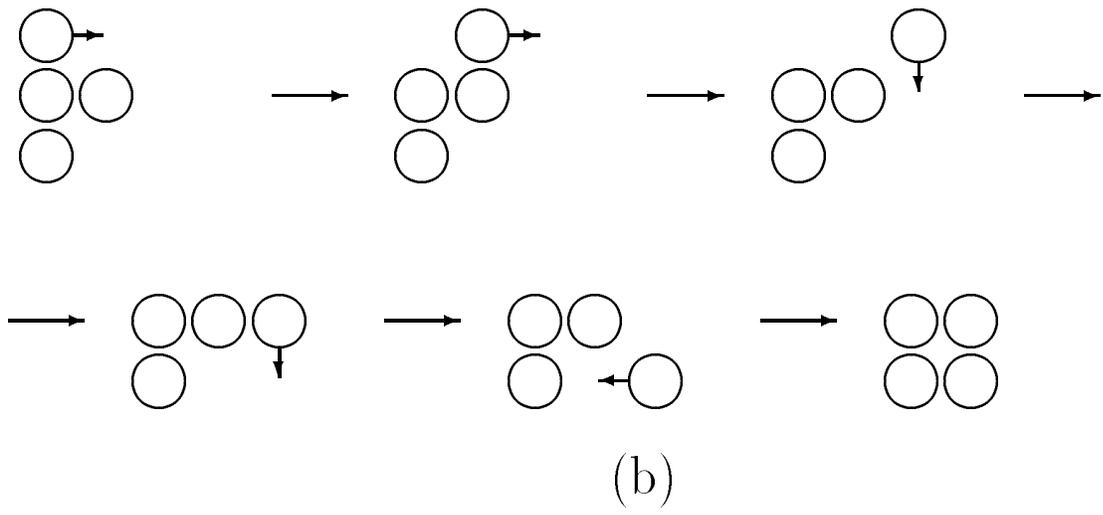
\begin{figure}
\caption{
The effect of allowing more moves on island morphology -
making islands more compact.
(a) the effect of allowing moves 2 and 32 on the structure 
of a six-atom island.
(b) the effect of allowing moves 3, 48, 6 and 96 on the structure
of a four-atom island.
}
\label{addmoves}
\end{figure}

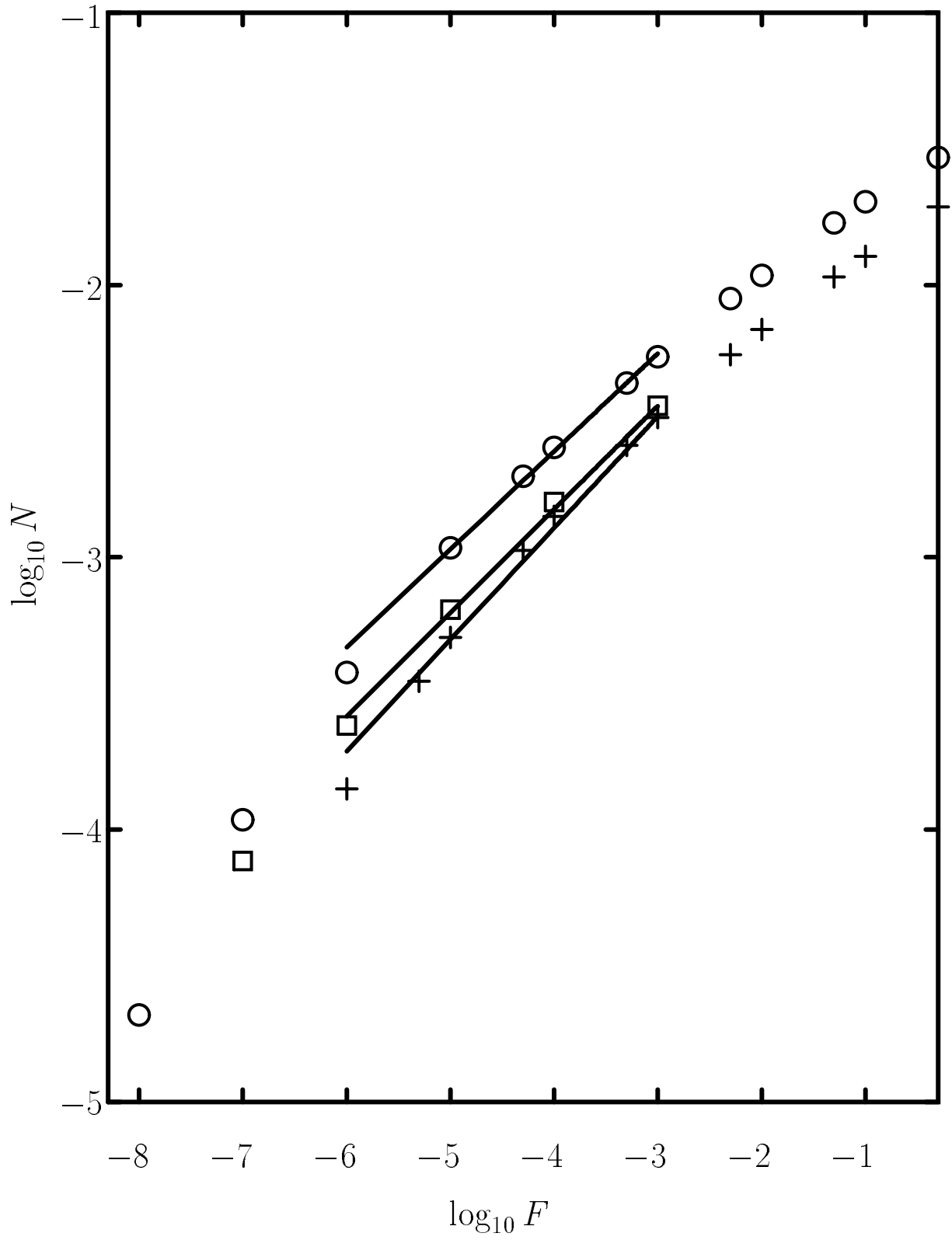
\begin{figure}
\caption{
The island density $N$ [islands/site] is plotted
vs. deposition rate $F$ $ML/s$
on a $\log-\log$ scale for
MC simulations of 
models I
($\bigcirc$),
II 
($\Box$)
and 
III 
(+)
(with $i^{\ast}=1,2$ and $3$
respectively)
at coverage of $\theta=0.2$
and substrate temperature $T=250 K$.
The exponent $\gamma$, given by the slope, 
increases as $i^{\ast}$ is increased. The results for 
$\gamma$ are summarized in Table
\ref{tab:scaling}
}
\label{scaling} 
\end{figure}

Note that the models 
studied here, 
and in particular models
II and III, are minimal in the sense that 
they include the minimal number of allowed moves for the
given $i^{\ast}$. One can add some more moves to each of these 
models without changing $i^{\ast}$. 
For example,
moves $n=$ 6 (96), 36 (66) and 70 (100) can be
added to model II.
These moves are obtained by taking the moves shown in Fig. 
\ref{moves} (b) 
and placing atoms in site $k=2$ or $k=6$ or both
(see Fig. \ref{local}).
The moves appear in pairs due to the horizontal mirror symmetry
such that the move in parenthesis is the mirror image of the one
which precedes it.
Similarly, moves 
$n=$ 7 (112), 52 (67), 70 (100) and 71 (116) can be added to model III. 
These additional moves have no significant effect on the scaling
properties (since $i^{\ast}$ remains unchanged)
but they enhance edge mobility making the islands
more compact. In particular, move 7 (112) allows atom mobility along
straight island edges which is not possible otherwise.
However, we find that
highly compact islands of nearly square shape observed on FCC(001)
metal substrates such as Cu(001) and Ni(001) cannot be obtained with
models which include only one hopping rate such as the models studied
here. Such highly compact islands are obtained only when 
the hopping rate for moves along island edges such as 7 (112)
are much larger than other hopping rates in the system,
including the hopping rate of the monomer
\cite{barkema94,amar95,biham96}.
In contrast, systems in which atoms can detach from island edges
and reattach elsewhere exhibit rather compact island shapes
even when the detachment rate is considerably smaller than
the hopping rate of the monomer
\cite{ratsch94,kandel95}.

In the models studied here diffusion occurs via the motion of
one atom at a time. Other diffusion mechanisms, which involve
concerted motion of two atoms have also been observed.
In particular, the
exchange move 
in which a substrate atom is displaced by an adatom and
pops out into a nnn site of the original adatom position.
On surfaces such as Al(001) 
\cite{feibelman90}
and Pt(001) 
\cite{kellogg90}
it was found that
the energy barrier for exchange is lower than for hopping
indicating that this is the dominant diffusion mechanism 
on these surfaces.
In a different type of concerted move, two atoms
move together along the edge of an island. When the barrier
for such move is lower than moves involving single atoms in
the island they may drive island diffusion
\cite{shi96}.
Recent STM experiments indicate that for Ag(001) even large islands
may diffuse as a result of the high edge mobility in these systems
\cite{wen94}. 

Models related to those studied in this paper
appear in the cluster-cluster aggregation problem,
where DLA-like clusters aggregate at a finite density
\cite{meakin,vicsek}.
Jensen {\it et. al.} 
\cite{jensen94}
have recently studied a model of island
growth in which edge mobility is suppressed while islands of
all sizes can move rigidly (and thus $i^{\ast}=\infty$)
while their diffusion coefficients decay according to
$h_s \sim 1/s^p$.
In this model the simple relation between
edge mobility and island mobility, which exists in 
the models of local interactions studied here is broken. 
Due to the lack of edge mobility islands maintain their DLA-like
shape while the exponent $\gamma$ which depends on $p$ is varied.

\section{Summary}
\label{sec:conclusions}

In summary, we have performed a systematic study of the effect of
mobility of small islands on the scaling and morphology in island
growth on surfaces. The exponent $\gamma$, which describes the
dependence of island density on deposition rate
was examined, 
for experimentally relevant deposition rates,
using both numerical integration of the rate 
equations and MC simulations. 
It was found that $\gamma$ 
increases as the critical island size $i^{\ast}$ is increased.
This reflects the decrease
in island density which results from the possibility of small islands
to move and merge. 
The asymptotic value of $\gamma$ in the limit of slow
deposition rate is given by
$\gamma = i^{\ast}/(2 i^{\ast} + 1)$.
However, convergence to this asymptotic value 
by decreasing the deposition rate
turns out to
be slow for MC simulations and even for rate equations.
In addition, $\gamma$ exhibits rather slow dependence
on $i^{\ast}$ as it increases and is limited to the range 
$1/3 \le \gamma \le 1/2$.
This indicates that using this scaling law to extract microscopic
information 
on mobility of small islands
from experimental results, is hard and requires very
precise measurement of $\gamma$.

The island morphology was found to  become more compact
as $i^{\ast}$ is increased.
This reflects a general 
relation between edge mobility and small island
mobility in models of short range interactions. 
To quantify the morphological changes we have performed a fractal
analysis of the island morphology using the box counting and mass
dimensions. 
In both cases it was found that the fractal dimension is rather 
insensitive to changes in
$i^{\ast}$, however the morphological
change is reflected in the value of the prefactors.

\acknowledgments

We would like to thank
I. Farbman, D. Kandel, D. Lidar (Hamburger) and O. Millo
for helpful discussions.

\newpage

\begin{table}
\caption{
Diffusion coefficients of small islands as measured
from MC simulations of single islands
for the three presented models at $T=250K$.
It is shown that $i^{\ast}=1,2$ and 
$3$ for models I,II and III respectively. 
}
\label{tab:hopping}
\end{table}

\begin{table}
\caption{
Comparison of $\gamma$ obtained from asymptotic analysis 
and numerical integration of rate equations, and from
MC simulations, for models I, II and III.
The numerical results 
for both the rate equations and MC simulations were obtained
at the same coverage $\theta=0.2$, and for the same
range of deposition rate $F=10^{-6}-10^{-2}$ $ML/s$ at $T=250K$.
}
\label{tab:scaling}
\end{table}

\begin{table}
\caption{
Box counting dimension $D_B$,
for models I, II and III at
three deposition rates $F$,
coverage $\theta=0.2$
and $T=250K$.
The dimensions are typically lower than the DLA dimension
and are only weakly dependent on the model.
}
\label{tab:d_b}
\end{table}

\begin{table}
\caption{
The  prefactor $A_B$ in Eq. 
(\ref{def:boxc_dim})
for the three models for various deposition rates $F$,
coverage $\theta=0.2$
and $T=250K$.
Entries are in units of $10^3$ boxes.
The prefactor is strongly dependent on the model reflecting
the morphological changes.
}
\label{tab:A_b}
\end{table}

\begin{table}
\caption{
Mass dimension $D_M$,
for models I, II and III at
various deposition rates $F$,
coverage $\theta=0.2$
and $T=250K$.
The dimensions are considerably higher than the
DLA dimension but only weakly 
dependent on the model.
}
\label{tab:d_m}
\end{table}

\begin{table}
\caption{
The prefactor $A_M$ in Eq. 
(\ref{def:boxc_dim})
for the three models for various deposition rates $F$,
coverage $\theta=0.2$
and $T=250K$.
The prefactor is strongly dependent on the model
and sensitive to the different morphologies.
}
\label{tab:A_m}
\end{table}

\newpage

\begin{center}
{\bf Table I}
\end{center}

\begin{table}
\vskip-\lastskip
\begin{tabular}{cr@{}l@{${}\pm{}$}r@{}l
		r@{}l@{${}\pm{}$}r@{}l
					r@{}l@{${}\pm{}$}r@{}l}
	Cluster & \multicolumn{12}{c}{Diffusion Coefficients
					[hops/s]}
	\\ 	\cline{2-13}
	Size & \multicolumn{4}{c}{model I} &
		\multicolumn{4}{c}{model II} &
		\multicolumn{4}{c}{model III} 
	\\	\hline

	1 & 82 &   & 2  &   & 82 &   & 2  &  & 84 &   & 2  & 
	\\
	2 & \multicolumn{4}{c}{0.0} & 21 &   & 1  &  &
		21 &   & 1  & 
	\\
	3 & \multicolumn{4}{c}{0.0} & \multicolumn{4}{c}{0.0} &
		7 & .7  &  & 0.5 
	\\
	4 & \multicolumn{4}{c}{0.0} & \multicolumn{4}{c}{0.0} &
		\multicolumn{4}{c}{0.0}
	\\
	\end{tabular}
\end{table}

\vspace{1.8in}

\begin{center}
{\bf Table II}
\end{center}

\begin{table}
	\begin{tabular}{ccr@{}l@{${}\pm{}$}r@{}lr@{}l@{${}\pm{}$}r@{}l}
	Model & \multicolumn{5}{c}{Rate Equations}
		& \multicolumn{4}{c}{Simulation}
	\\	\cline{2-6}
	& \multicolumn{1}{c}{Asymptotic} & \multicolumn{4}{c}{Numerical}
		& \multicolumn{4}{c}{}
	\\	\hline

	I & 1/3=0.333 & 0 & .31 & 0. & 01 & 0 & .36 & 0 & .01
	\\
	II & 2/5=0.400 & 0 & .36 & 0. & 01 & 0 & .38 & 0 & .01
	\\
	III & 3/7=0.430 & 0 & .37 & 0. & 01 & 0 & .41 & 0 & .01
	\\
	\end{tabular}
\end{table}

\newpage

\begin{center}
{\bf Table III}
\end{center}

\begin{table}
	\vskip-\lastskip
	\begin{tabular}{cr@{}l@{${}\pm{}$}r@{}l
				r@{}l@{${}\pm{}$}r@{}l
					r@{}l@{${}\pm{}$}r@{}l}
	Model & \multicolumn{4}{c}{$F=10^{-6}$ [ML/s]}
		& \multicolumn{4}{c}{$F=10^{-7}$ [ML/s]}
		  & \multicolumn{4}{c}{$F=10^{-8}$ [ML/s]}
	\\	\hline

	I & 1 & .62  & 0 & .01 & 1 & .66 & 0 & .01 & 1 & .68 & 0 & .02
	\\
	II & 1 & .59 & 0 & .01 & 1 & .64 & 0 & .01 & 1 & .66 & 0 & .02
	\\
	III & 1 & .59 & 0 & .01 & 1 & .67 & 0 & .01 & 1 & .72 & 0 & .02
	\\
	\end{tabular}
\end{table}

\vspace{1.8in}

\begin{center}
{\bf Table IV}
\end{center}

\begin{table}
	\vskip-\lastskip
	\begin{tabular}{cr@{}l@{${}\pm{}$}r@{}l
				r@{}l@{${}\pm{}$}r@{}l
					r@{}l@{${}\pm{}$}r@{}l}
	Model & \multicolumn{4}{c}{$F=10^{-6}$ [ML/s]}
		& \multicolumn{4}{c}{$F=10^{-7}$ [ML/s]}
		  & \multicolumn{4}{c}{$F=10^{-8}$ [ML/s]}
	\\	\hline

	I & 18 & .7 & 0 & .5 & 19 & .4 & 0 & .5 & 20 & .2 & 0 & .5
	\\
	II & 14 & .7 & 0 & .5 & 15 & .6 & 0 & .5 & 15 & .9 & 0 & .5
	\\
	III & 10 & .9 & 0 & .5 & 11 & .9 & 0 & .5 & 12 & .0 & 0 & .5
	\\
	\end{tabular}
\end{table}

\newpage

\begin{center}
{\bf Table V}
\end{center}

\begin{table}
	\vskip-\lastskip
	\begin{tabular}{cr@{}l@{${}\pm{}$}r@{}l
				r@{}l@{${}\pm{}$}r@{}l
					r@{}l@{${}\pm{}$}r@{}l}
	Model & \multicolumn{4}{c}{$F=10^{-6}$ [ML/s]}
		& \multicolumn{4}{c}{$F=10^{-7}$ [ML/s]}
		  & \multicolumn{4}{c}{$F=10^{-8}$ [ML/s]}
	\\	\hline

	I & 1 & .84 & 0 & .02 & 1 & .89 & 0 & .02 & 1 & .83 & 0 & .02
	\\
	II & 1 & .89 & 0 & .02 & 1 & .94 & 0 & .02 & 1 & .83 & 0 & .02
	\\
	III & 1 & .91 & 0 & .02 & 1 & .90 & 0 & .02 & 1 & .88 & 0 & .02
	\\
	\end{tabular}
\end{table}

\vspace{1.8in}

\begin{center}
{\bf Table VI}
\end{center}

\begin{table}
	\vskip-\lastskip
	\begin{tabular}{cr@{}l@{${}\pm{}$}r@{}l
				r@{}l@{${}\pm{}$}r@{}l
					r@{}l@{${}\pm{}$}r@{}l}
	Model & \multicolumn{4}{c}{$F=10^{-6}$ [ML/s]}
		& \multicolumn{4}{c}{$F=10^{-7}$ [ML/s]}
		  & \multicolumn{4}{c}{$F=10^{-8}$ [ML/s]}
	\\	\hline

	I & 2 & .1 & 0 & .1 & 1 & .9 & 0 & .1 & 2 & .3 & 0 & .1
	\\
	II & 2 & .5 & 0 & .1 & 2 & .0 & 0 & .1 & 2 & .9 & 0 & .1
	\\
	III & 3 & .3 & 0 & .1 & 3 & .1 & 0 & .1 & 3 & .5 & 0 & .1
	\\
	\end{tabular}
\end{table}

\newpage

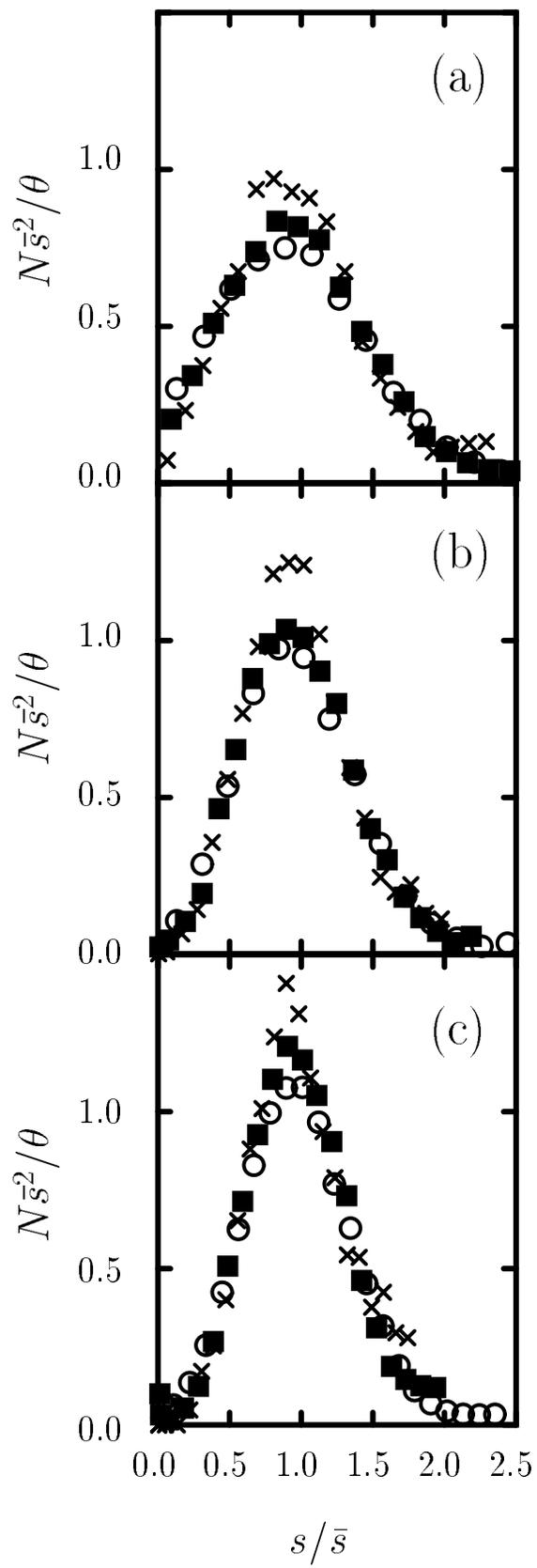
\begin{figure}
\caption{
Scaled island size distribution vs. scaled size for model
I (a), model II (b) and model III (c). The deposition rates
are 
$F=10^{-3}$ $ML/s$ ($\bigcirc$),
$F=10^{-4}$ $ML/s$ 
(\vbox{\hrule width 8pt height 8pt})
and
$F=10^{-5}$ $ML/s$ ($\times$)
and the coverage is
$\theta = 0.2$.
Note that the peak narrows as $i^{\ast}$ increases.
}
\label{histogram} 
\end{figure}

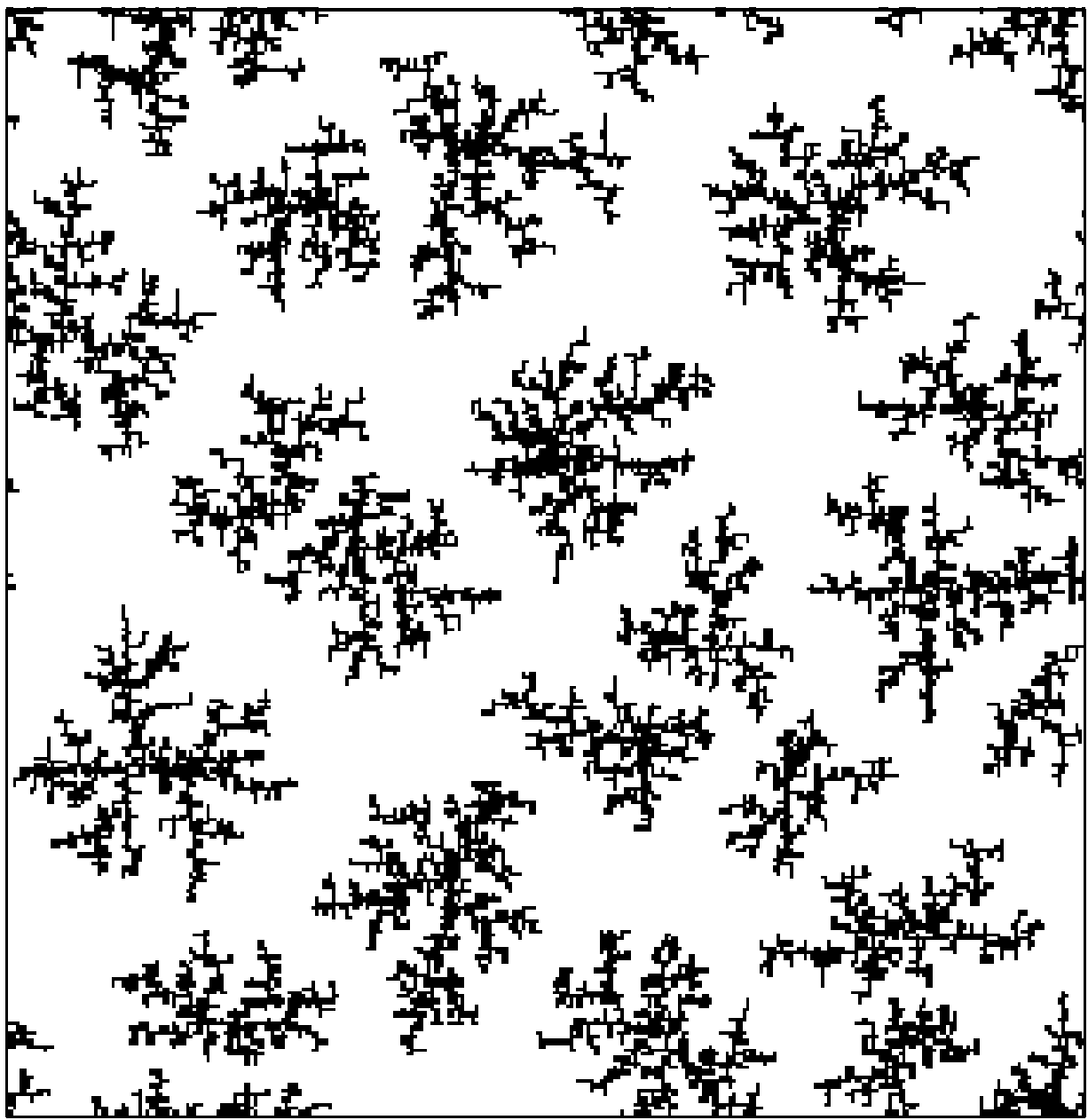
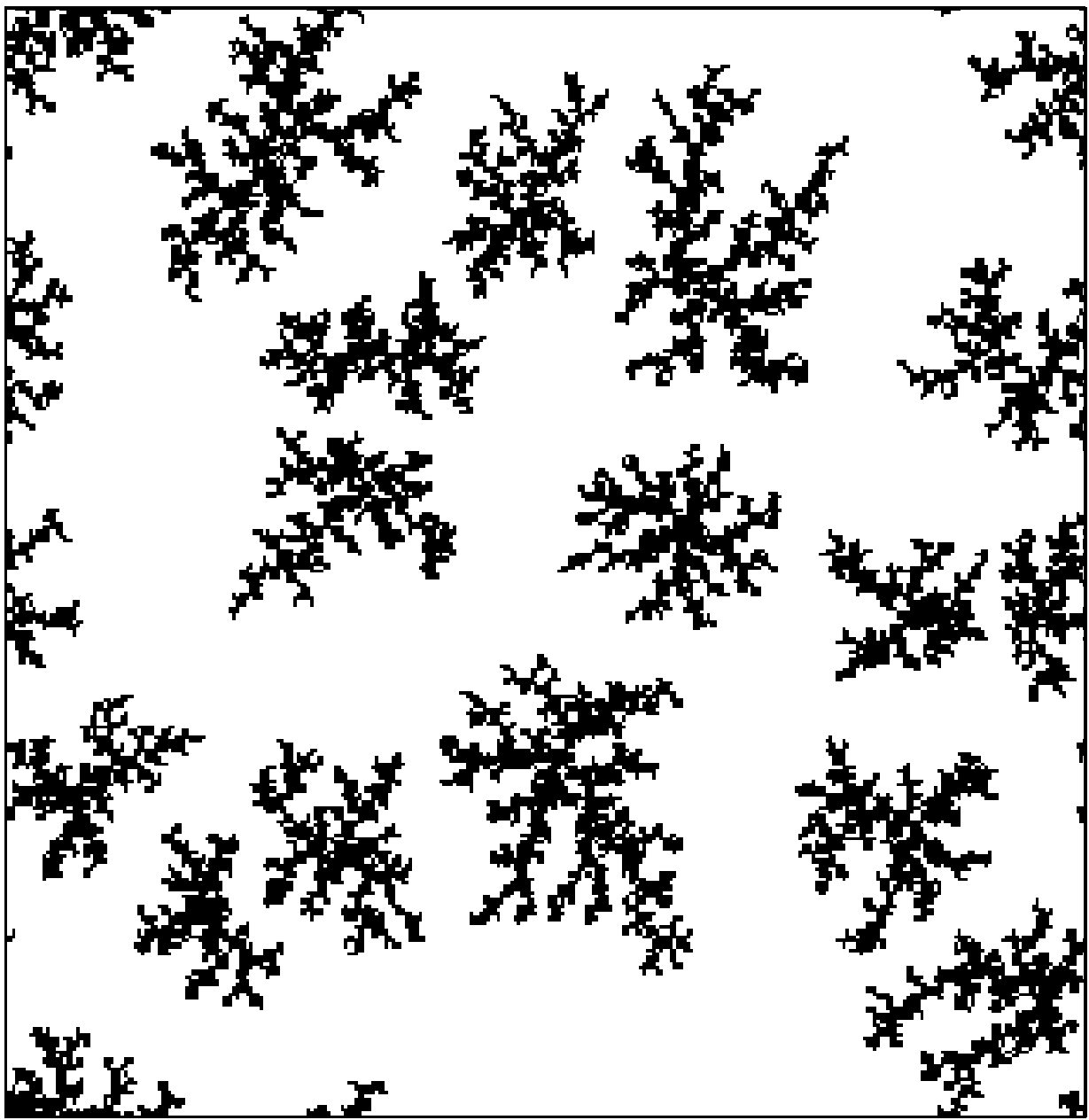
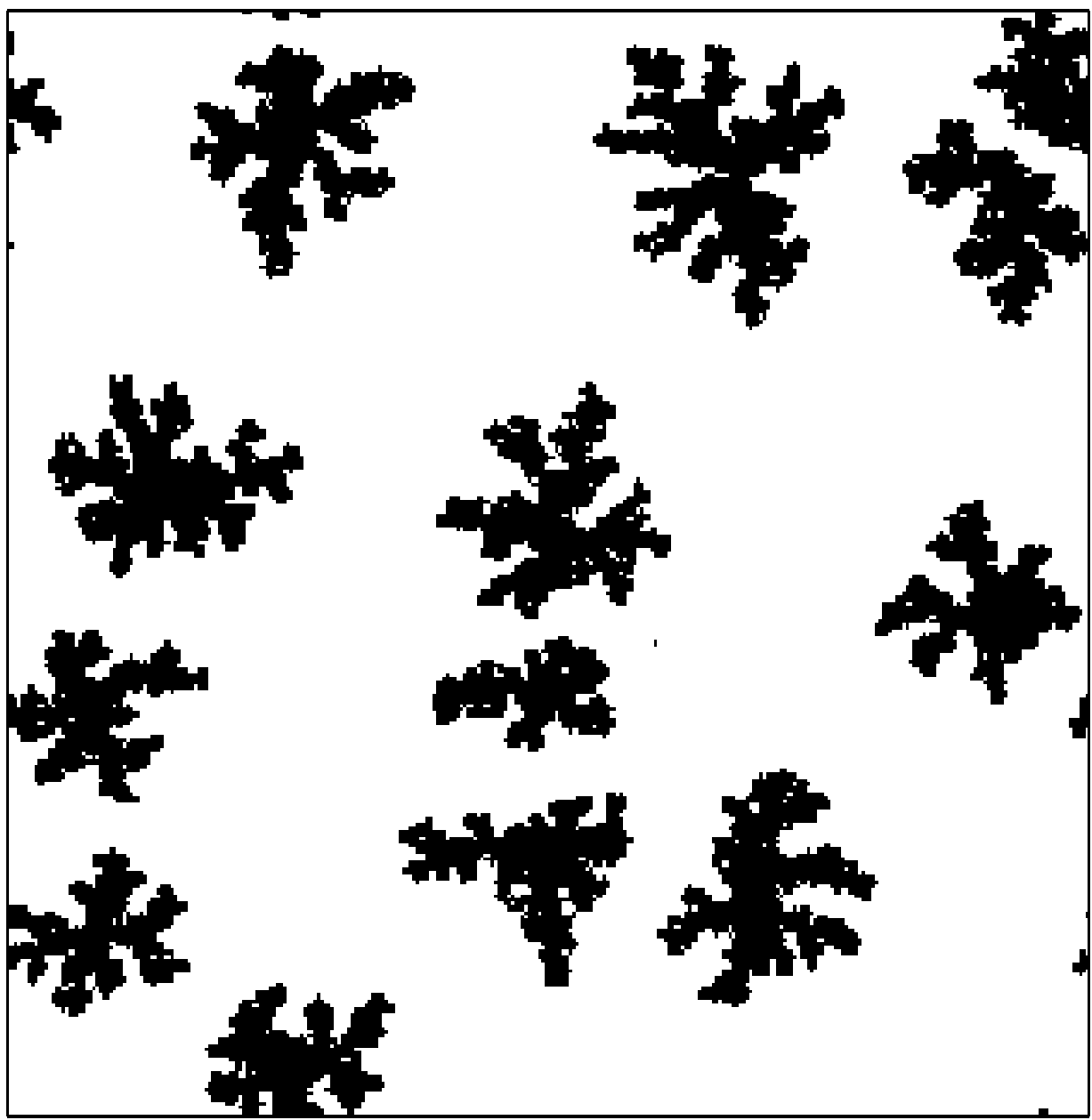
\begin{figure}
\caption{
Top view of surface layer under growth conditions specified by
Models I (a), II (b) and III (c)
for deposition rate of $F=10^{-6}$ $ML/s$,
coverage $\theta=0.2$
and substrate temperature $T=250 K$.
The deposited atoms and islands are represented by the dark
color and the exposed substrate is white.
Lattice size: $250 \times 250$ sites.
}
\label{snaps}
\end{figure}

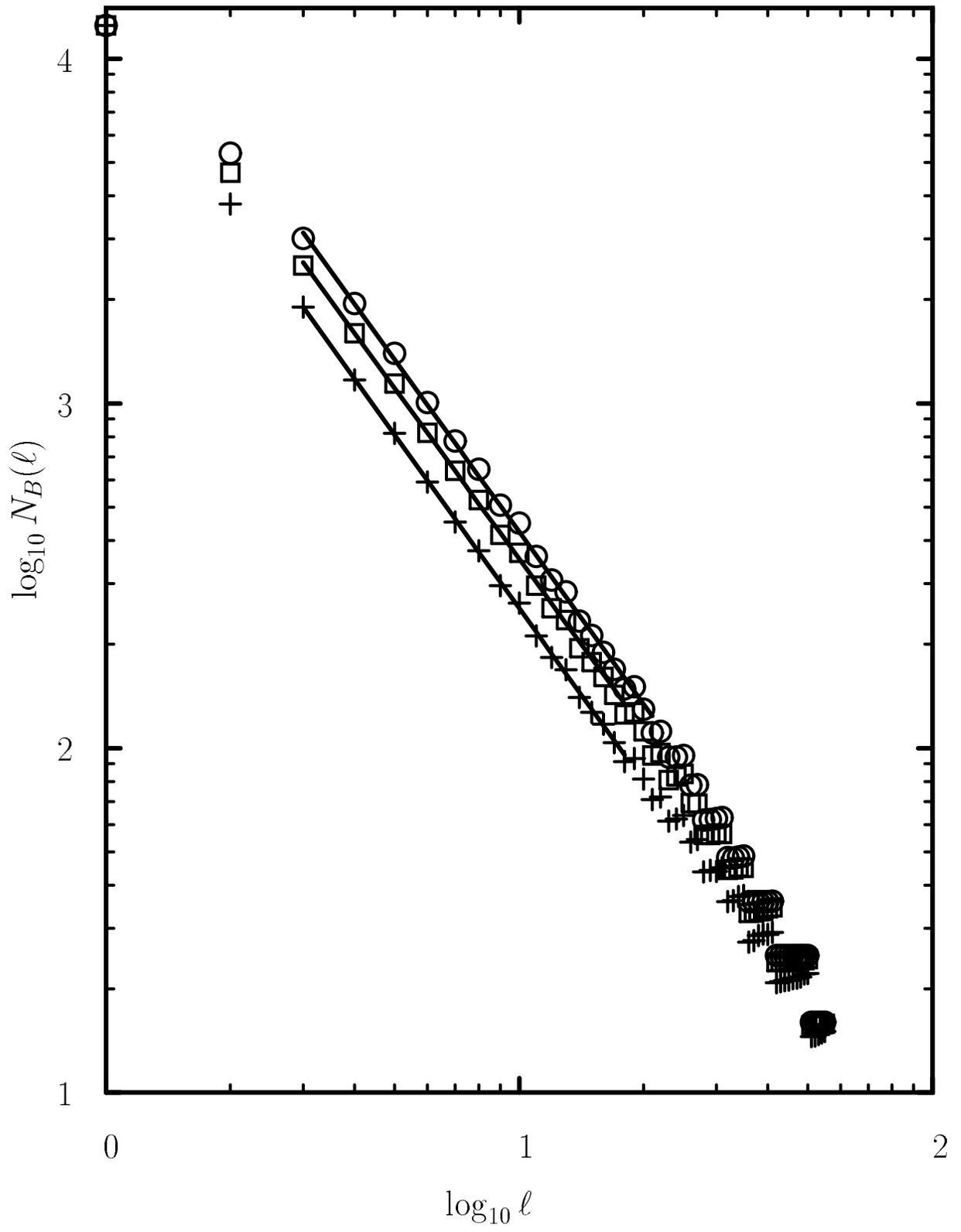
\begin{figure}
\caption{
The box-counting function which counts the number $N_{B}(\ell)$ of
occupied boxes vs. box size $\ell$ is shown for the island
morphology obtained from 
models 
I ($\bigcirc$), 
II ($\Box$)
and 
III (+)
at deposition rate $F=10^{-7}$ $ML/s$,
coverage $\theta=0.2$
and substrate temperature $T=250 K$.
Note that the slopes are found to be
the same for the three models indicating that the fractal
dimension $D_B$
in insensitive to the small island
mobility
(Table III). 
However, the intercept 
(determined by the prefactor $A_B$)
is different 
for the three models (Table IV).
}
\label{boxc} 
\end{figure}

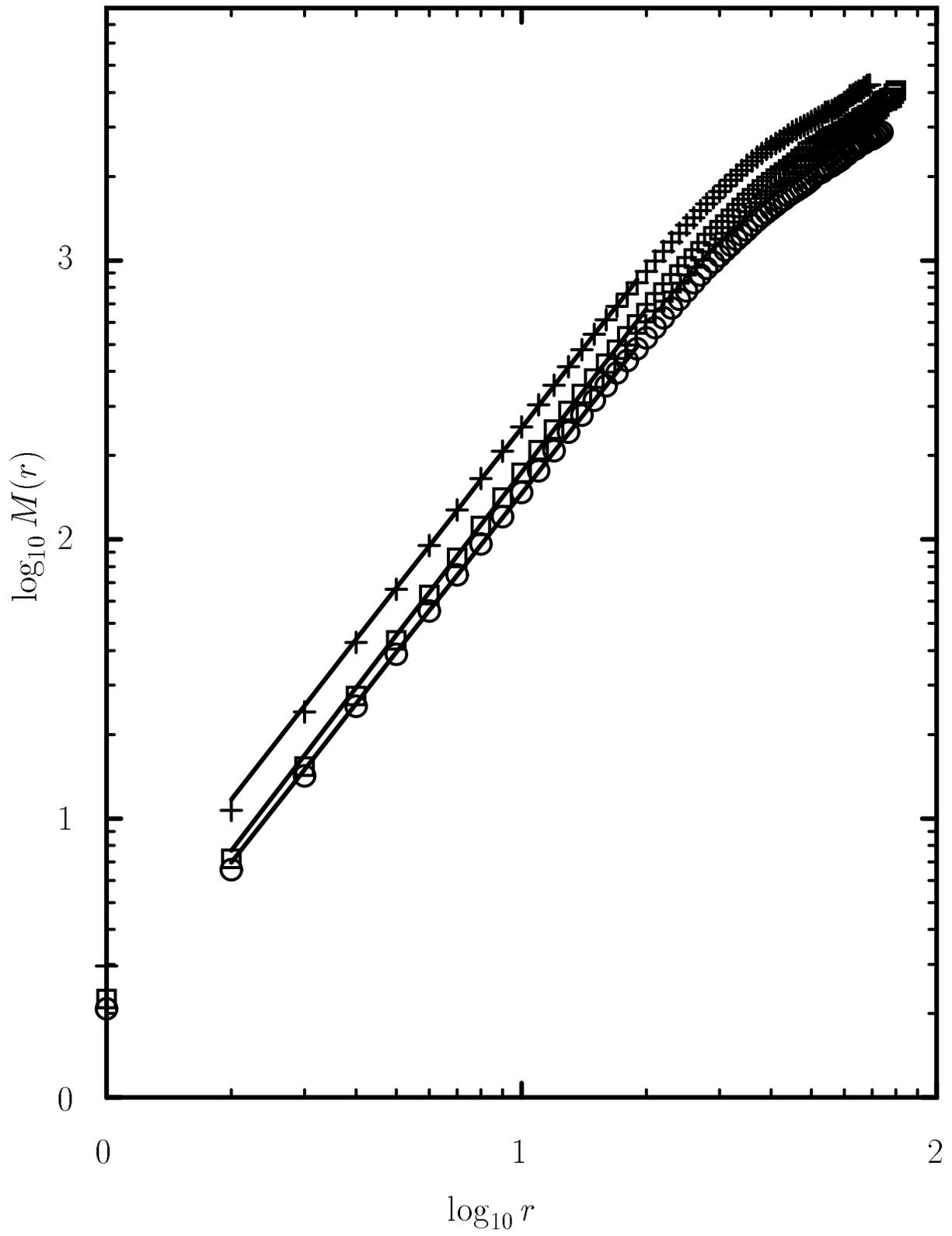
\begin{figure}
\caption{
the mass $M(r)$ within a circle of radius $r$ around the 
center of mass of an island vs. $r$ for islands obtained
from models 
I ($\bigcirc$), 
II ($\Box$)
and 
III (+)
at deposition rate $F=10^{-7}$ $ML/s$,
coverage $\theta=0.2$
and substrate temperature $T=250 K$.
The mass dimension is given by the slope on the $\log-\log$ scale.
It is shown that the
mass dimension $D_M$
is insensitive to the different morphologies
obtained from these models (Table V). 
However, the prefactor 
$A_M$
shows 
significant differences (Table VI).
}
\label{massd} 
\end{figure}


\begin{references}
\bibitem{diffusion_coef}
The diffusion coefficient for a monomer is given by
Eq. 
(\ref{hop_rate})
with $E_B=0.5$ $eV$ and $T=250 K$. The result is
$h_1 = 84$ $hops/s$.
It can be shown that the dimer's diffusion coefficient $h_2$
satisfies $h_2=h_1/4$.
\bibitem{oldscalingrelation}
The scaling relation between $N$ and $F/h$ is in agreement
with the result previously obtained in Ref. 
\cite{villain92}
where a different approach was used.
\end{references}
\end{document}